# PARTICIPATORY DESIGN:

# A SYSTEMATIC REVIEW AND INSIGHTS FOR FUTURE PRACTICE

Peter Wacnik, Shanna Daly, Aditi Verma[1]
University of Michigan

## ABSTRACT

Participatory Design – an iterative, flexible design process that closely involves stakeholders, often end users – is growing in use across design disciplines. As more practitioners use Participatory Design (PD), it has become less rigidly defined, with stakeholders engaged to varying degrees through disjointed techniques. This ambiguity can be counterproductive when discussing PD processes. We performed a systematic literature review that builds shared, foundational knowledge of PD processes and techniques while also summarizing the state of PD research in the field, as a first step in supporting richer understandings of how best to equitably engage with stakeholders. We found that a majority of PD literature examined specific case studies of PD, with the design of intangible systems representing the most common design context. Stakeholders most often participated throughout multiple stages of a design process, recruited in a variety of ways, and engaged in several of the 14 specific participatory techniques identified. Our findings also identify leverage points for creators of PD processes and how the leverage points impact design equity, including: (1) emergent vs. predetermined processes; (2) direct vs indirect participation; (3) early vs late participation; (4) one time vs iterative participation; and (5) singular vs multiple PD techniques.

## 1. INTRODUCTION

Participatory Design (PD) is a design approach aimed at developing technologies with close involvement from stakeholders – especially those most affected by the result, often end users. Participatory Design typically involves multiple rounds of requirements gathering, prototype development, implementation, and evaluation (Hardie, 1988). Originating in Scandinavian countries in the 1970s, PD was initially used to empower unions with action-oriented design methodologies (Bjögvinsson et al., 2012). One such instance involved the Norwegian Iron and Metal Workers Union (NJMF), where union representatives worked with government researchers to investigate new technologies for the workplace (Ehn, 1988). The NJMF research project resulted in multiple proposals for more efficient computer-based, shop-floor planning systems, changes in work organization on the shop floor, and a textbook compiled to educate union workers on planning, control, and data processing in their work.

[1]additive@umich.edu

However, over time, the use of PD has evolved, becoming less rigidly defined as a specific process or used in a specific context. Instead, it has become an overarching term encompassing projects that engage stakeholders in multiple ways at various stages and kinds of design work. Various methods for involving stakeholders in design – such as inclusive design, user-centered, human-centered, co-design, customer co-creation, and crowdsourcing – are all considered participatory within this broader framework (Aitamurto et al., 2015). However, these terms are often used interchangeably, even when stakeholders are not consistently or directly involved in the project. This inconsistency contributes to the vague definition of Participatory Design, leading to conflicting interpretations and gaps in practitioners' understanding of the concept. Rather than asserting what PD is or is not, our work highlights core characteristics of how people have applied what they refer to as PD processes and summarizes the current state of PD research. Our goal is to establish a foundation for future studies that can enhance understanding and promote the broader adoption of meaningful Participatory Design processes. We reviewed 88 design articles that discussed applications of Participatory Design from seven academic journals and five conference proceedings focused on design. Examining the literature, our review specifically focused on the types of research, design contexts, timing of participation, strategy of participation, applied techniques, and recruitment methods.

There is growing recognition that inequities arise from improper design practices, and a strong desire across design disciplines – particularly in engineering – to address these inequities by engaging directly with stakeholders and users. However, conflicting interpretations of Participatory Design and gaps in understanding successful PD processes inhibit the achievement of equitable design. Misunderstood or poorly designed PD processes may even exacerbate inequities. For PD to contribute to equitable design, practitioners must have a deep and shared understanding of PD processes. This literature review analyzes past applications of PD, establishes a foundational understanding of contextualized PD processes, and identifies research gaps, discussing learnings necessary to further successful PD.

## 2. BACKGROUND

Participatory Design has evolved significantly since its inception in the late 1970s in Scandinavia. Design thinking itself can be categorized as a modern interpretation of PD, with emphasis on the need for designers to address the social implications of innovation, collaborate with a diverse set of stakeholders throughout the process, and develop multiple prototypes to examine potential ideas for their effectiveness (Bjögvinsson et al., 2012). Bjögvinsson et al. demonstrate how PD's core values, including democratic and direct user participation and acknowledging participants' tacit knowledge, were pivotal in shifting designers' mindsets from designing objects to designing 'socio-material assemblies' involving stakeholders.

Scholars have since considered similar principles in the contexts of their design work. For example, Winschiers-Theophilus et al. investigated experiences in rural African communities, noting that it is widely accepted among designers that user involvement in a design process leads to better outcomes for the stakeholders, but that user involvement has been variable across projects (Winschiers-Theophilus et al., 2012). They argued for a deeper exploration of the meaning of participation in design and its potential impact on design outcomes, particularly in cross-cultural contexts. They further claim that achieving meaningful participation requires mutual learning among designers and local community members, and that a variety of methods exist to facilitate that process, emphasizing that designers must gain in-depth local knowledge to guide the choice and adaptation of participatory methods.

Synthetic reviews of Participatory Design have been completed by other researchers as well. One such review finds that current definitions for Participatory Design are too narrow and lacking, leading to inconsistencies in PD processes that negatively affect the research and advancement of such design practices (Aitamurto et al., 2015). These researchers call for a more comprehensive understanding of these design processes – an understanding that can begin to be achieved through a broad survey of literature and an analysis of PD approaches.

## 3. METHODS

We conducted a systematic literature review on how Participatory Design has been researched and practiced. This review was guided by three research questions

1. What are the foundational characteristics and techniques of Participatory Design that span different contexts, design processes, and stakeholder groups?
2. What is the current state of Participatory Design research in the field?
3. How can future research dive deeper into gaps in understanding to create a fuller picture of modern Participatory Design?

### 3.1. Literature Search

The literature search was conducted from journals defined by Gemser et al. as top design journals as well as proceedings from popular design conferences (Gemser et al., 2012). The journals and proceedings included in the literature search are listed in Table 1 below.

Table 1. Sources Included in the Literature Search

| **Academic Journals** | **Conference Proceedings** |
| --- | --- |
| *AI EDAM*<br><br>*Design Science Journal*<br><br>*Design Studies* | *American Society of Mechanical Engineering Design Theory and Methodology Conference*<br><br>*Communications of the Association for Computing Machinery* |
| *Design Issues* | *Computer Supported Cooperative Work* |
| *Journal of Mechanical Design* | *International Conference on Engineering Design* |
| *Research in Engineering Design* | *The Design Conference* |

We used the query “Participatory Design” within the title, abstract, or keywords to identify relevant articles. This search produced 151 articles that were filtered to remove duplicates and articles that were not full-length journal publications. This filtering process narrowed the literature to 95 items. We then excluded review papers of PD, resulting in 88 articles for analysis. This process is represented in Figure 1, and includes the number of articles from each of the journals or conference proceedings we reviewed – the publication cutoff for inclusion was May 2022.

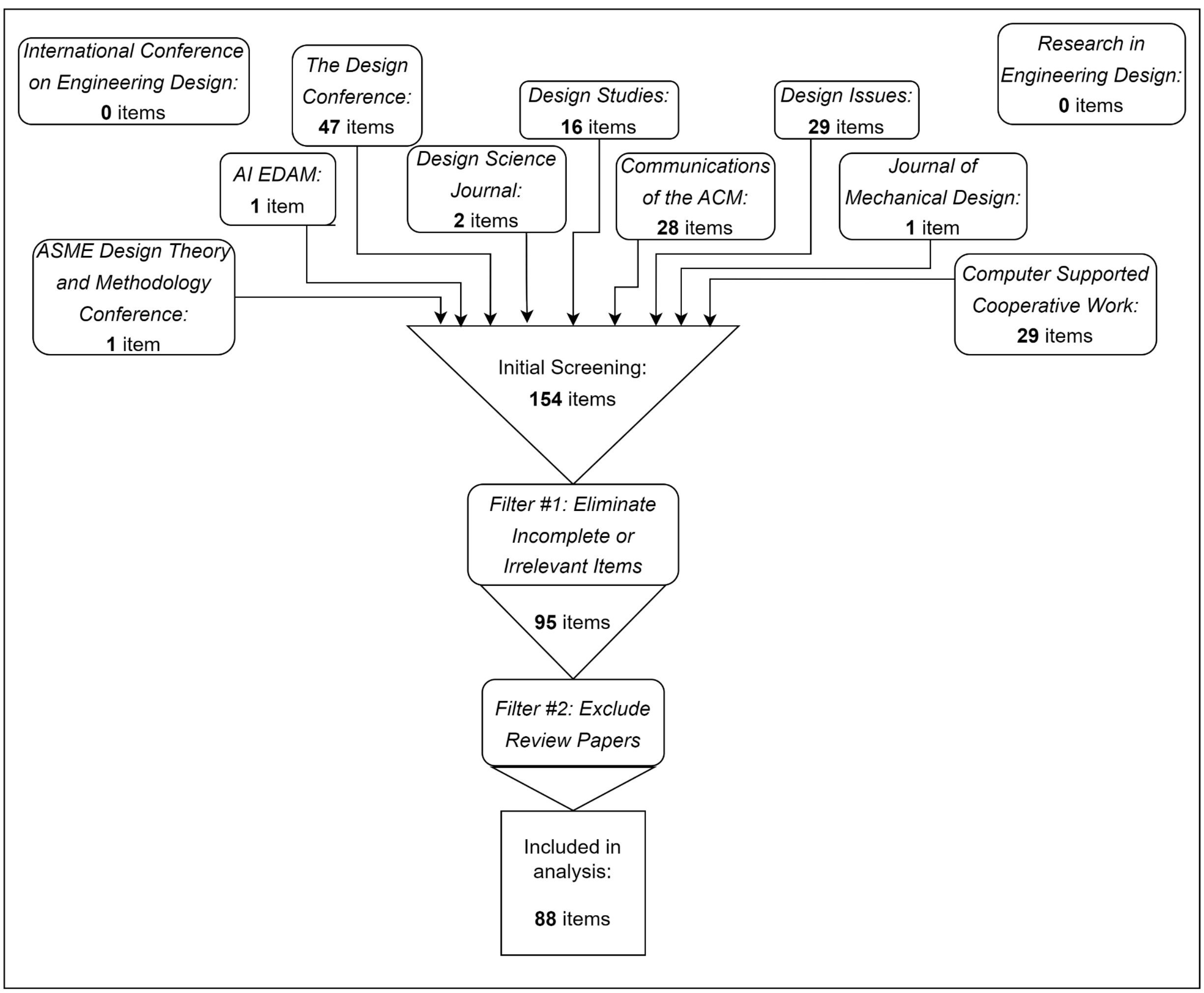

Fig. 1. Filtering process to determine peer-reviewed articles to include in analysis.

### 3.2. Analysis

Two researchers began the literature analysis by operationalizing the research questions previously listed into analysis categories of PD characteristics to better guide the review of articles: Type of Participatory Design Research; Context of Design; Stakeholder Recruitment; Timing of Participation; Participatory Techniques; and Strategy of Stakeholder Participation. Each analysis category was developed over time through regular reviews and discussion with the research team as trends began to emerge and new information was gathered from the literature. The final codebook is listed in Table 2.

Table 2. Codebook for PD Characteristics Represented in Literature

| *Article Analysis Category* | *Definition* |
| --- | --- |
| Type of Participatory Design Research | General classification to sort the articles, describing the type and scope of PD covered in each article: (1) Participatory Design Process Applications; (2) Participatory Design Technique Analysis; (3) Participatory Design Principles; (4) Guidelines for Participatory Design; and (5) Reflections on Participatory Design. |
| Context of Design | A description of what was being designed and the intended outcome of the design process. Initially classified into four high-level categories to describe the context in which PD was discussed or applied: (1) Artifacts; (2) Intangible Systems; (3) Physical Systems; and (4) Design Process Critiques.<br><br>Additionally distinguishes the environment described by the literature as a real world project or a theoretical experiment/reflection. |
| Stakeholder Recruitment | Specific information and techniques that the author used to recruit participants for PD activities. This includes methods of outreach, location of activities, participation incentives, and recruitment efficacy, if discussed. |
| Timing of Participation | Categorizes the timing of stakeholder participation into one of four general stages of the design process: (1) Front End; (2) Middle End; (3) Back End; (4) Throughout.<br><br>Additionally distinguishes one-time stakeholder participation from iterative stakeholder participation. |
| Participatory Techniques | Identifies the 14 specific techniques and methods that were used in the PD process discussed in each item of literature. |
| Strategy of Stakeholder Input | Identifies the techniques used by the authors as predetermined (a set plan to involve stakeholders) or emergent/changing (a flexible process that adapted with the stakeholders and design changes). |

Once the categories in the analysis matrix had been established, the two researchers reviewed entries of the matrix for each other, swapping six articles to cross-check the review process and matrix data. This approach provided a way of checking reliability and ensured that all of the data collected across the items of literature was consistent and therefore suitable for analysis.

## 4. FINDINGS

### 4.1. Types of Participatory Design Research

We found five Types of Participatory Design Research, described in Table 3: (1) Participatory Design Process Applications; (2) Participatory Design Technique Analysis; (3) Participatory Design Principles; (4) Guidelines for Participatory Design; and (5) Reflections on Participatory Design.

Table 3. Types of Participatory Design Research; Classifications and Definitions

| *Type of Participatory Design Research (Of 88 articles)* | *Definition* | *Example* |
|---|---|---|
| Participatory Design Process Applications<br><br>53 articles; 60% | Articles that examined a full PD process in a specific context to determine the effectiveness of a participatory approach. | A case study of the participation of disadvantaged women in Hong Kong, in a design process for the purposes of affecting government policy (Kwok, 2004). |
| Participatory Design Technique Analysis<br><br>13 articles; 15% | Articles that focused on a specific technique for fostering participation and involving stakeholders in the design process. | A study investigating the effectiveness of three-dimensional models to foster stakeholder input and participation in Botswana to discover resident preferences for street infrastructure and home design (Hardie, 1988). |
| Guidelines for Participatory Design<br><br>9 articles; 10% | Articles that encompassed directives for PD within a specific context, offering a prescriptive evaluation of how to effectively use PD. | An article that described infrastructuring techniques beyond the initial stages of PD through a case study introducing new fabrication technologies to a Danish school system, accompanied by tenets to guide others in implementing their expanded technique (Bødker et al., 2017). |
| Participatory Design Principles<br><br>9 articles; 10% | Articles focused on fundamental, overarching elements or characteristics of PD independent of design context. | An investigation of the "mundane and strategic" work that permeates a Participatory Design process, such as coordinating workshop space, finding participants, or scheduling the timing of activities (Hyysalo & Hyysalo, 2018). |
| Reflections on Participatory Design<br><br>4 articles; 5% | Articles that critiqued PD experiences from a practitioner's perspective, including specific successes or failures. | An article describing pitfalls in the prototype testing experience in the development of an Electronic Health Record prototype, prompted by attempts to rectify dissimilar stakeholder needs (Bossen, 2006). |

### 4.2. Context of Design

A majority of the articles discussed projects, processes, or case studies in real-world contexts (81 articles; 92%) while a small number examined theoretical discussions or reflections of Participatory Design (7 articles; 8%). Beyond the general division of real-world versus theoretical studies, the specific contexts in which PD was applied are as follows:

- *The design of Artifacts* (6 articles; 7%) – PD used to support the design of products used by one or a small group of end users, whether intended for consumer sale or other uses. One subcategory within this group was the design of accessible technology. For example, one article described the creation of two different accessible devices: an intelligent mobility aid for the elderly to navigate crowded areas and alleviate stress from crowds, and an active wheelchair for athletic users (Wilkinson & De Angeli, 2014).

- *The design of Intangible Systems* (61 articles; 69%) – PD was used to support the design of software or other non-physical systems. This category included activity design, workflow management, and organizational processes. As an example, one article in this category (subcategory electronic information management) described a PD project focused on the system for document preservation for brittle books in university libraries (Anderson & Crocca, 1993). Another example article, categorized as a public sector project, discussed a PD project to empower new-arrival women to Hong Kong, to have a voice in the government processes and policy surrounding housing and urban planning (Kwok, 2004).

- *The design of Physical Systems* (12 articles; 14%) – Articles discussed the use of PD to design physical systems to be used by a large group of many end users, such as buildings, urban planning projects, and workspace design, as opposed to the fewer end users of artifacts. An example of an urban planning project was an article that investigated new purposes for an obsolete railway track in Belgium by building community narratives with extensive resident participation (Huybrechts et al., 2018).

- *Design Process Critiques* (9 articles; 10%) – Some articles were contextualized in designers' experiences with PD processes or techniques, rather than the output of a specific project, and focused on reflections and evaluations of PD processes or techniques. For example, one article evaluated a role-playing game participatory approach, where peers interacted with each other and the game to share experiences with the New York welfare system (Campbell, 2004).

Figure 2 illustrates subcategories within these broader categories, noting the number of articles in each.

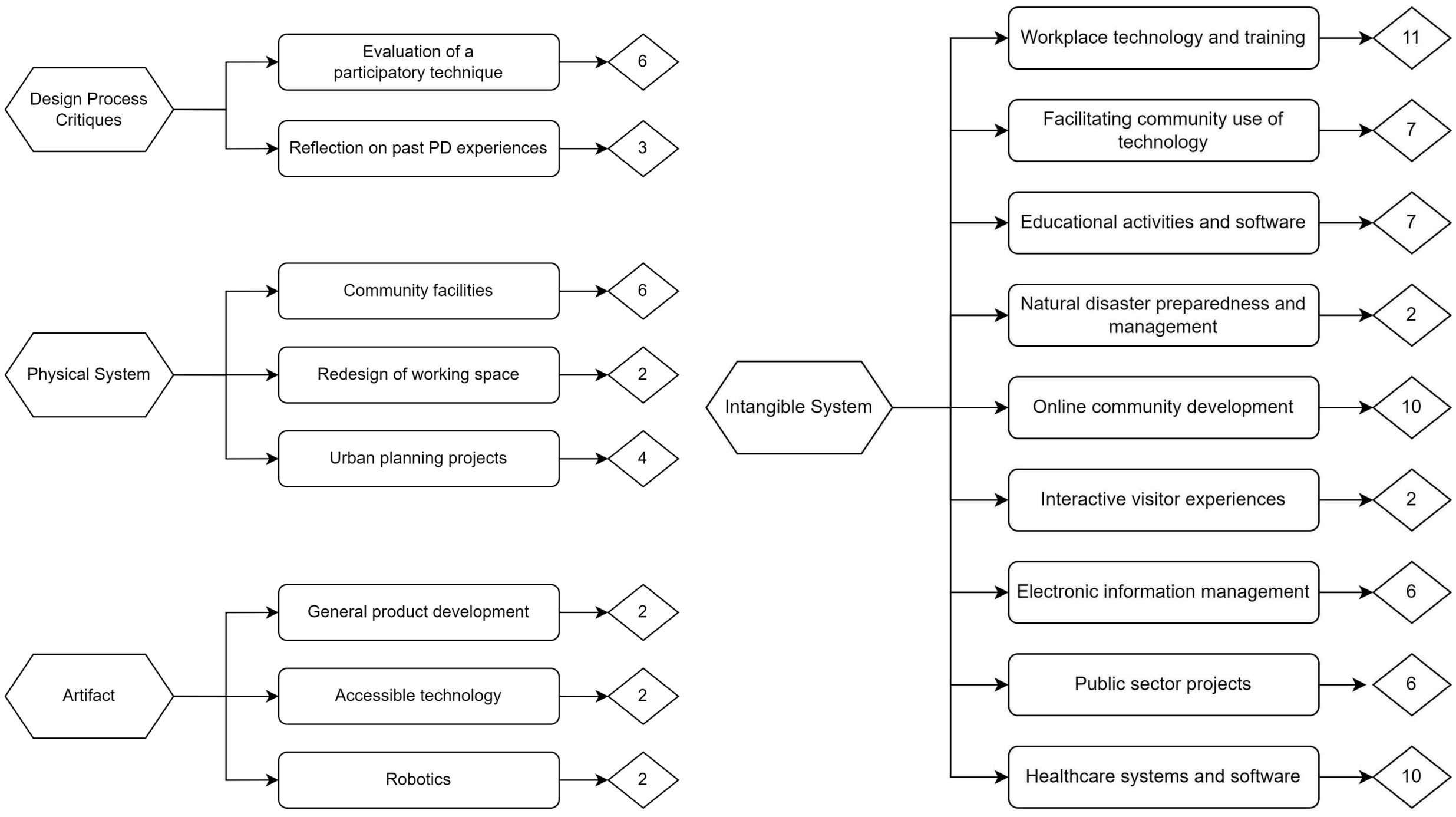

Fig. 2. Design contexts in which PD was discussed (article count in each subcategory).

### 4.3. Stakeholder Recruitment

Even if a design process is participatory, it is not necessarily equitable – effective and equitable stakeholder recruitment methods are a key first step to ensuring that a participatory process is set up to be equitable. The majority of papers that discussed stakeholder recruitment focused on the characteristics of the participant pools (62 articles; 70%) when commenting on recruitment methods. The six types of participant recruitment approaches discussed are described in Table 4 below.

Table 4. Methods of Stakeholder Recruitment for Participation

| *Stakeholder Pool*<br>*(Of 88 articles)* | *Recruitment Methods* |
|---|---|
| Target stakeholder demographic<br><br>(17 articles; 19%) | Practitioners developed a target stakeholder demographic and proactively recruited participants from that pool. Designers reached out over email, through workshops, through personal or professional networks, or with posters. Some projects randomly selected participants from a target stakeholder pool to contact. |
| Expert users by application<br><br>(4 articles; 5%) | Participants were members of social media groups or practitioners' networks and were offered the opportunity to apply to participate through those channels. In other cases, peers or organization administrators identified participants as experts. |
| Open to the public<br><br>(9 articles; 10%) | Practitioners invited communities to participate through participatory events and workshops held in public spaces (physical and online), posters, word-of-mouth, or fliers handed out by the research team. |
| Volunteers in interested organizations<br><br>(10 articles; 11%) | Practitioners identified or were contacted by interested organizations and recruited people within those organizations. |
| Employees from a stakeholder company<br><br>(14 articles; 16%) | The design work involved a specific company and employees from that company were recruited to participate. Employees either volunteered for the project or were directed by management to participate. |
| Students from a class<br><br>(8 articles; 9%) | Seen in educational contexts, students were contacted to participate through emails, announcements to the class, or directed to participate. |

Most often, designers identified a target stakeholder demographic that they believed would bring the most useful insights to the design process, or be the most affected by the design outcome. Once this population was identified, designers reached out over email, through workshops, through their networks, or posters. In one example, practitioners designing a memory aid for people with amnesia recruited multiple amnestics along with a rehabilitation specialist and computer scientist through their professional networks (Wu et al., 2004). Some projects randomly selected participants from a target stakeholder pool to contact for participation – this method was mostly seen when designers used surveys as a participatory technique.

In cases where practitioners had little to no previous knowledge of the design context, they relied on stakeholders to bring a depth of understanding to the design process, sometimes co-designing the solution, which necessitated expert users as participants. Practitioners identified these experts through communications with their networks, their peers, or stakeholder organizations before offering them a chance to apply to join the design team and subsequently selecting expert participants for the project. For example, in the design of new product opportunities for the athletic wheelchair user market, practitioners recruited four Paralympians to be expert users through their network – their status as experts sufficiently proven – to be lead users involved throughout the design process (Wilkinson & De Angeli, 2014).

The most open recruitment approaches involved design activities that were open to the public. For example, in an article describing the development of a community library, recruitment was very open and allowed all citizens the chance to participate (Dalsgaard, 2012). Practitioners accomplished this by leveraging participatory techniques in the library that invited people to record feedback as they walked by. Another example with open public recruitment saw researchers place posters in busy public areas, hand out fliers, and rely on word-of-mouth to reach stakeholders (van Manen et al., 2015).

Engaging volunteers in an interested organization, employees at a company, and students from a class involved similar recruiting methods for designers. Commonly, these organizations or classes had specifically requested a project that utilized Participatory Design, and participants volunteered due to their awareness of the project or were directed by respective management to engage with designers. In cases where participants were not directed to engage, they learned of the design process through emails, posters, or announcements made by their organization. Two such examples of this type of recruitment saw hospital staff engaged in the development of digitized X-ray examination technology (Kjær & Madsen, 1995) and students that redesigned educational activities (Guha et al., 2005) – participants were aware and involved due to their investment in the outcome and being directed by higher-level authority.

### 4.4. Timing of Participation

We summarize the stages at which articles described stakeholder participation in Figure 3. The timing of stakeholder engagement was not discussed in 7 of the 88 articles.

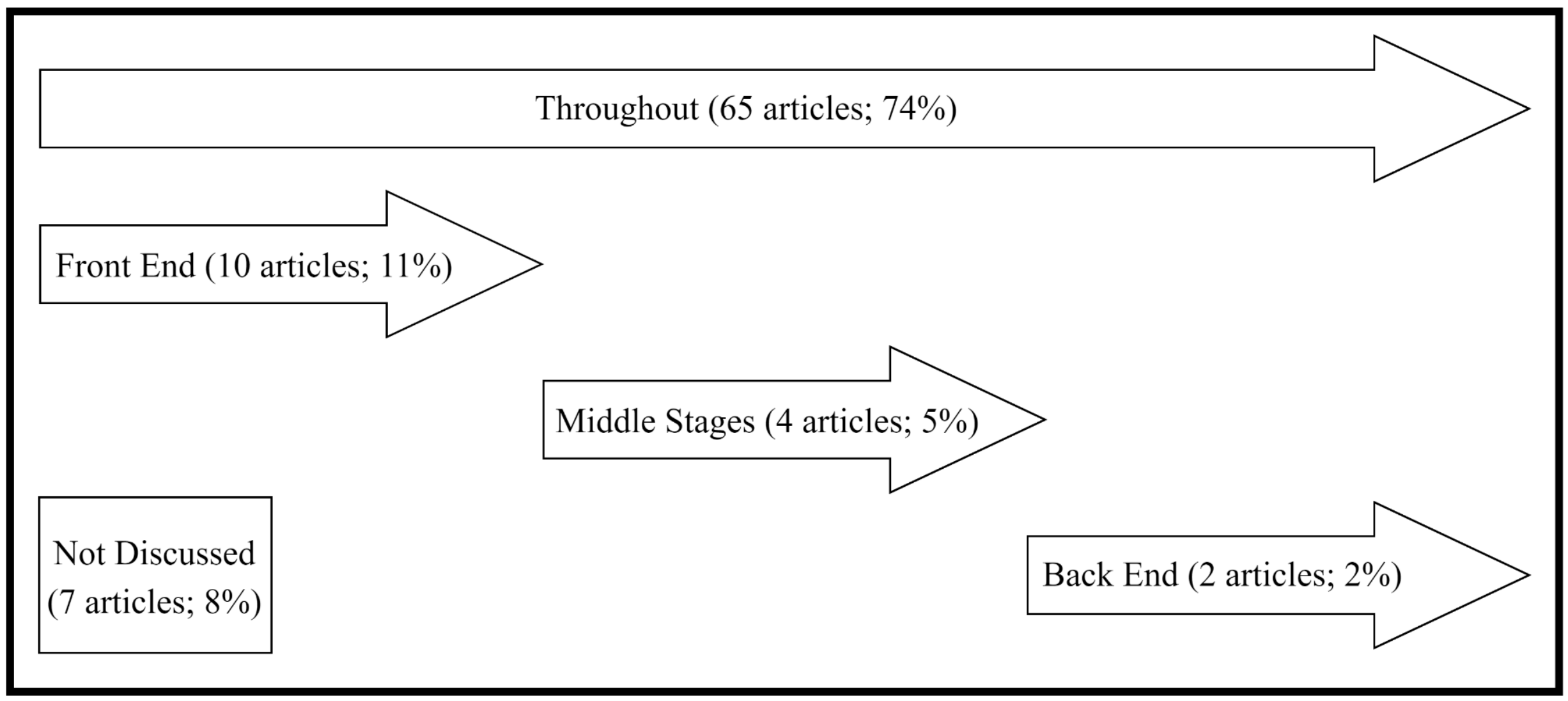

Fig. 3. Timing of stakeholder participation in the articles quantified.

Most articles (65 articles; 74%) described stakeholder participation throughout a design process at multiple stages of the work. For example, interaction design researchers developing interactive technologies for a municipal library involved stakeholders through reflections on the importance of the library, discussion of their visions for a future building, cogeneration and evaluation of design concepts for interactive technologies, and a plan to continue community involvement through the remainder of the project (the article was written before the new library had been constructed) (Dalsgaard, 2012). From the very beginning and throughout the project, the practitioners maintained stakeholder participation as a guiding principle for their work, articulating in the project's core values that stakeholder participation would be the foundation on which design decisions were made.

Some articles (10 articles; 11%) specifically sought participation from stakeholders in the front end of a design process. For example, one article described a project repurposing an old coal track and the community participation – participatory workshops, interviews, prototyping, and context-specific activities – occurred during the early stages of the work that lasted 16 months (Huybrechts et al., 2018). At its conclusion, the project team had crafted multiple alternative uses for the track using input from the community.

A handful of articles (4 articles; 5%) described stakeholder participation in only the middle stages of a design process. In one example, at a project for a new university in southern Sweden to design workspaces, users were involved mainly during the prototype evaluation phase with VR technologies, testing and providing feedback on four prototypes the design team had developed (Davies, 2004).

The fewest number of articles (2 articles; 3%) involved stakeholders at only the back end of a design process. Tapped In, an online community aimed at supporting education professionals, used methods at the back end of the design process to sustain the infrastructure previously built for the community (Farooq et al., 2007). Users were asked to specifically contribute in developing the infrastructure by providing consistent feedback once the first iteration of the system had been implemented.

### 4.5. Frequency of Participation

Six papers (7%) described one time participation, *single instances of engagement*, where the designer utilized one participatory interaction at one stage of a design process. In an article where designers investigated solutions to increase self-reliance during volcanic disasters in Costa Rica, the research team held two participatory workshops on consecutive days – one participatory interaction at one stage of the design process – that involved a questionnaire, individual and group brainstorming, and initial concept filtering and prioritization of needs (van Manen et al., 2015). The designers took insights from these one-off workshops to apply to the design process.

The other 79 articles (90%) described iterative participation, *engaging participants in multiple activities* within or across front-, middle-, and back-end activities of a design process. One example saw a research team explore speech-based operation of computers during dental surgeries (Cederman-Haysom & Brereton, 2006). The research had an iterative process that began with ethnographic studies with a large number of dentists and dental students, before identifying three specialists that participated in techniques such as low fidelity prototypes, design games, and role-playing throughout the design process. There were also three one-on-one design sessions with these specialists resulting in a prototype that was evaluated through discussion and a pilot trial during an operation. Three of the 88 articles (3%) did not discuss the frequency of stakeholder involvement.

### 4.6. Participatory Techniques

We identified 14 unique participatory techniques described in the articles, as methods to facilitate participation from stakeholders in the design work. Nearly all of the articles, 79 of 88 (90%) used multiple techniques for stakeholder participation. The full list of techniques are described in Table 5, along with an example of each technique.

Table 5. Specific Participatory Techniques Identified in Literature

| *Technique* | |
|---|---|
| *Description* | *Example* |
| Participatory Workshops – (64 articles; 73%) | |
| Designers and participants met together in a mutual learning situation for input by stakeholders, learning about the design context, ideating solutions together, or evaluating the design path. Workshop activities included futures workshops, scenarios, ideation, design games, concept evaluation, problem or solution mapping, and stakeholder reflections. | Using co-ideation to develop more self-reliance in the face of volcanic disasters in Costa Rica, two, two-hour participatory ideation workshops were held in two central places near the volcano (van Manen et al., 2015). First, participants were given a questionnaire to gain initial insights. Then, they ideated on Post-its, subsequently collating Post-its into central themes and illustrating ideas. Each group selected one idea to develop further, presented their idea, and all groups voted to select their favorite. |
| Stakeholder Interviews – (61 articles; 69%) | |
| Interviews were conducted to gain a deep understanding of the stakeholder. Designers used semi-structured interviews, unstructured conversations, or user-led visits that enabled both an interview along with a demonstration of their user experience. | Investigating the implementation of computer support for the Editorial Board of a Film Board to streamline their workflow, interviews were conducted with multiple people from multiple stakeholder groups, with follow-up interviews as well (Simonsen & Kensing, 1997). During the dialogue, the authors viewed how participants completed tasks and heard their design suggestions. This built mutual learning situations between designers and stakeholders, which resulted in drawings of the current workflow and potential improvements to the system. |
| Prototyping with Stakeholders – (52 articles; 59%) | |
| Prototypes were presented to stakeholders or stakeholders were asked to build prototypes themselves. Stakeholders were able to visualize the solution and how it might be embedded into the relevant context. This included low-fidelity or high-fidelity prototype builds, stakeholders evaluating or reacting to prototypes, or a pilot installation of a prototype. In some instances, prototyping occurred during participatory workshops or prototype evaluation during interviews – in these cases, we counted the activity as a workshop or interview respectively, in addition to prototyping. | In the development of speech and gesture technology to be used during dental surgeries, researchers performed an ethnographic study with dentists and dental students, followed by multiple workshops with a demonstration of an existing low-fidelity prototype to elicit feedback (Cederman-Haysom & Brereton, 2006). A functional higher-fidelity prototype was developed and trialed with a dentist in-practice, gaining critical insights for designers. |

| *Technique* | |
| --- | --- |
| *Description* | *Example* |
| Context-Specific Activities – (19 articles; 22%) | |
| Context-specific activities were novel techniques developed by designers to engage stakeholders in a particular, unique context – with the activity likely not transferable to other projects. Most often, this involved notably modifying a participatory technique to better-suit the unique design situation. Some types of context-specific activities in the literature included a unique design game simulating a welfare system, guiding children through observation, or open-to-the-public displays to record stakeholder feedback, among others. | During the early stages of a project to motivate families to monitor power consumption and reduce electricity spending, the authors invented and facilitated an at-home card game for participating families to reflect on their power consumption practices (Albrechtslund & Ryberg, 2011). Doing so allowed families to ease into the PD process with a context-specific technique later leading to additional participation . |
| Update Meetings with Stakeholders – (13 articles; 15%) | |
| Update meetings were held with stakeholders to share progress reports and information. Designers presented this information to stakeholders and took questions or feedback. | To align with new reforms from the Danish Ministry of Education that emphasized 21st century skills in the classroom, a design team worked with three Danish municipalities to embed digital fabrication technology and design thinking into lower secondary schools with a hybrid learning space (Bødker et al., 2017). A steering committee was formed that met with the municipalities on a quarterly basis to discuss progress, share results, and receive feedback. These update meetings kept stakeholders up to date and informed of the progress. |
| Committee of User Representatives – (14 articles; 16%) | |
| User representatives acted as a type of committee to represent a larger stakeholder group. Rather than recruit participants for each event, practitioners leveraged this committee of the same user representatives – sometimes expert users – for most (if not all) of the participatory activities. As the user representatives participated in most other activities during the design process, we counted this technique in addition to other techniques used to engage the user representatives, such as workshops or focus groups. | During the design of a new, online entertainment system, designers utilized a Wiki forum to communicate with users (Hess & Pipek, 2012). In addition, they developed a larger user parliament of day-to-day users and a central committee that was composed of elected, expert users and staff members. With this two-group user representation, designers took input from a wide range of users in the parliament while meeting with the central committee weekly to make design decisions and implement functionalities. |

| *Technique* | |
| --- | --- |
| *Description* | *Example* |
| Focus Groups – (10 articles; 11%) | |
| Focus groups created an environment that was conducive to more insights and consensus built from different perspectives within the stakeholder group. Designers included community members in focus groups most commonly, to discuss stakeholders' lifestyles and relevant thoughts about the design context. At times, stakeholders also evaluated prototypes in focus groups. | The early stages of a design project for the UK PM involved normally excluded citizens in focus group discussions to elicit feedback on the concept of an access token system for personal identification and admittance to public services (Dearden et al., 2006). They began with an introduction and open discussion of participants' lifestyle issues with public services and utilities. They then discussed a more theoretical topic of smartcards or other media to assist the citizens with their lifestyle difficulties to help prompt insightful discussion. |
| Public Hearings – (9 articles; 10%) | |
| In public hearings, designers presented design paths and the process to be followed, explicitly organized to garner feedback from the public. It was also inherent that these hearings were open to the public for feedback from any stakeholder who feels they have insights to contribute. These were commonly seen in projects that dealt with large community infrastructure development. | In the development of a building to house a municipal library and the Citizens' Service Department in Aarhus, Denmark, the team was tasked with designing and integrating new, interactive technologies and services into the building (Dalsgaard, 2012). They leveraged public hearings where aspects of the building were presented and the floor opened to discussion with stakeholders once the information was conveyed. |
| Stakeholder Observations – (43 articles; 49%) | |
| Observations were leveraged by designers to get a firsthand view of the stakeholder's life, sometimes to the point of experiencing daily life with them. This appeared as observing natural tendencies at home or workflows of stakeholders in an organization, separated from the lifestyle, as well as embedding themselves in the design context, experiencing the environment that stakeholders do every day. Documentation methods of observations included written notes, pictures, or recordings. | Investigating the implementation of a new hardware/software system for digitized X-rays in a new hospital building for a radiology department, researchers utilized multiple sessions of observation (Kjær & Madsen, 1995). These were conducted at various locations in the hospital – secretary workplaces, during meetings within the department, in examination rooms – while taking pictures to document the workplace before implementing the new system. The authors emphasized recording the changes from the new system, and building on the department's regular work. |
| Stakeholder Surveys – (17 articles; 19%) | |
| Designers used surveys to obtain a large sample size of insights from stakeholder groups. With well-developed questions, designers gleaned quantitative data to survey the state of the design context and learn about stakeholders. | At the Institut Pasteur in Paris, a design team sought to create software tools to support scientific databases and network infrastructure (Letondal & Mackay, 2004). The authors conducted a campus-wide survey during the early stages of the design process that included 40 questions across various |

| *Technique* | |
|---|---|
| *Description* | *Example* |
| They also collected qualitative data in open-ended questions on the survey, learning about stakeholders' lifestyles on a deeper level. | categories of software use and needs, garnering 600 responses. The findings mapped the different stakeholder groups at the institute, providing the researchers more contextual information. |
| Competitive Benchmarking – (6 articles; 7%) | |
| In competitive benchmarking, designers created a survey of the current problem space while also identifying opportunities for new innovations. This technique took many forms, including a review of academic literature to gain an understanding of similar research contexts or applications, or benchmarking to evaluate current solutions from competitors and understand the gaps, guiding improvements for the future outcome. | During the design process for Sprock-it – a "hand-sized robotic character that encourages full-body interaction and engaging mental play" for children – the design team began by benchmarking competitor toys and devices (Burleson et al., 2007, p. 1). Stakeholders were indirectly involved in the benchmarking, as designers took four of the most popular, analogous products to benchmark – including the stakeholder's voice based on the popularity, without consulting them directly. The designers analyzed the functionalities of each of these devices and how they accomplished the desired user experience. |
| Historical Document and Data Analysis – (13 articles; 15%) | |
| Historical document and data analysis involved reviewing internal documents and historical data to obtain an overview of the stakeholder organization, organizational workflow, or design context. Designers coordinated with stakeholders to obtain the most relevant documents and data, before separately analyzing it to build a foundational overview of the design context. This commonly occurred at the beginning of the participatory design process to get designers up to speed. | During an investigation of the construction and maintenance of a wireless community network (WCN) in Italy called Ninux.org, two authors began their process with a document review (Crabu & Magaudda, 2018). They included local reports, articles, and other materials with a focus on methods of communication for users. This review led to a discussion of themes regarding the WCN, which informed the author's initial understanding of the problem landscape and contextual data. |
| Infrastructuring for Continued Participation – (9 articles; 10%) | |
| Infrastructuring is a technique particularly unique to PD, aimed at building a system for stakeholder independence at the conclusion of the design process. Designers engaged stakeholders in a series of meetings, organizational changes, and a hand-off process to ensure seamless implementation of a solution and sustainable development by stakeholders into the future. This occurred with both virtual systems or physical systems. | About 200 members of a freelancer network that lived and worked throughout Germany used a program named SIGMA that provided them with technical equipment and software (Törpel et al., 2003). An infrastructuring method allowed for a continuous design process undertaken by the freelancers. The members, over time, built a strong foundation of system knowledge, using past experience to develop the system and bring new users into the continuous design process. |

| *Technique* | |
|---|---|
| *Description* | *Example* |
| Stakeholder Personas or Scenarios – (7 articles; 8%) | |
| Designers used personas or scenarios as a form of indirect stakeholder participation during the design process. Practitioners developed personas or scenarios using previous stakeholder insights to represent an imaginary stakeholder or a common use-case situation respectively. After making these profiles, designers referenced and reflected on them consistently throughout the remainder of the design process. | Due to roadblocks in policy, the designers of OutBurst – a child-centric, online environment for children to react to and express their emotions about current events – were not able to bring children to the studio during the design process (Antle, 2004). Instead, they developed a series of personas to indirectly bring children into the design process while they worked. The designers brainstormed multiple personas and eventually used one named Rachel and a second named Dodge. Designers consistently referred back to how Rachel and Dodge would think or feel about various design decisions to help guide the process and outcome. |

### 4.7. Strategy of Stakeholder Input

The strategy of stakeholder input used throughout the process was classified as either *predetermined* or *emergent* to investigate the level of flexibility and dynamics of stakeholder agency. For *predetermined* Participatory Design processes, we defined the category as a PD process in which the techniques are pre-planned, followed specific guidelines for execution, and generally did not deviate from this initial plan; 39 articles (44%) used participatory techniques in this way. In one such example, researchers developed reading software for kids to make it more interactive and engaging (Kaplan et al., 2006). The research team first conducted a contextual inquiry about children's reading habits before running a preliminary study where children used reading software to read a book for four weeks. Researchers observed their use, took data, and held meetings to discuss the children's experiences.

*Emergent* Participatory Design processes had more nuance in the process execution. While the techniques were identified beforehand, the overarching goals and execution of the techniques actively evolved as the practitioners managed the design process; 47 articles (54%) used participatory techniques in this way. Emergent design processes included unplanned iterations on techniques – for example, circling back to a specific stakeholder group with additional interviews at a later stage in the design process to glean additional insights (Ginige et al., 2014). In another example, a researcher aiming to improve the wastewater management systems of low-income communities in Indonesia began with interviews to identify existing concerns amongst the public (Rosenqvist, 2018). This approach included a context-specific design game developed and played with participants to collaboratively evaluate, reflect, and iterate on the responsibility of stakeholders in wastewater management. Another design game was played with an expanded set of stakeholders to allow for further discussion with interviews conducted after the workshops to identify any shifts in matters of concern. Two articles did not discuss the dynamics of participatory techniques in the design process.

## 5. DISCUSSION

Looking at trends in the findings more broadly provides a shared foundation of knowledge of PD processes and techniques that have been leveraged across a variety of contexts. These trends help summarize the current state of PD in design science, revealing gaps where further investigation is needed to understand the full impact of all characteristics and variables in a Participatory Design process.

### 5.1. Types of Participatory Design Research Trends

Of the five types of Participatory Design, a majority (53 articles; 60%) discussed specific case studies of PD within a design project. This type of research is critical for communicating design projects to the academic community and records the successes and challenges of a specific design process in a specific context. Each design process and context is different, with particularities and nuances that are not found in past work – a PD process to build a community library in Denmark (Dalsgaard, 2012) is significantly and justifiably different from the process to develop educational software to engage children in active reading (Kaplan et al., 2006). In this specific comparison, practitioners investigating the reading software would have needed to translate the specifics from the PD process for the library into general guidance, rather than being able to pull from a higher-level, foundational document outlining Participatory Design best practices, such as our literature review. The next largest share of the literature, 15% of articles evaluated a Participatory Design technique – a category that is similarly limited in scope to the case studies, as it does not evaluate the larger picture of PD. One such example delved into the nuances of 3D models to allow residents of a Botswana settlement to model their own houses and discover the preferences of residents as to the street patterns of new areas in the settlement, with the research focusing on the effectiveness of the 3D models in discovering stakeholder needs (Hardie, 1988). This article and similar others are more specific than case studies, looking deeper at one particular technique or activity in a Participatory Design process – again, while extremely useful for practitioners to understand the nuances of a technique, this type of articles does not take the principles of Participatory Design and generalize them for different design contexts.

The remaining 25% of articles reviewed fit best into categories discussing guidelines applicable to a certain PD context, foundational principles for any PD process, or an author's reflections on the efficacy of PD in design. More often than not, these papers drew the guidelines or principles from a few design studies and missed aspects of PD from peer research. We believe this gap is notable in the Participatory Design research landscape, as articles that examine guidelines or principles might be directly applied to future projects. Such articles may also examine the deeper concerns of inequitable design – guidelines for a specific design context aimed at empowering a historically disadvantaged stakeholder group, or overarching principles that enable a design process foundationally imbued with equity, empowerment, and mutual learning. We did not find articles that specifically addressed the gaps of PD guidelines and principles for equitable design, but similar works could have a positive impact and align with our analysis in this paper.

### 5.2. Design Context Trends

A key metric showed the heavy majority of articles – 81 articles, or 92% – discussed real-world applications of PD. The findings of nearly all articles used in this paper were not the conjecture of researchers. Rather, real projects with real stakeholders produced the learnings in this paper, indicating the likely success of these learnings if applied to future, real-world projects.

The more granular classification of Design Contexts in PD displays a prevalence of intangible systems – 61 of 88 articles (69%). This aligns with the historical origins of PD in the 'Scandinavian Approach' from the 60s and 70s that developed from industries involving trade unions with the design and implementation of workplace systems and processes (Farrell et al., 2006). Although PD has also been historically used in the architecture discipline as well (Davies, 2004), the disparity of physical system contexts to intangible systems is interesting and shows that PD may be underutilized in the design of physical systems. Closing the identified gap across disciplines and contexts may help PD become more accessible and utilized across a wider breadth of design projects.

**5.3. Consistent Stakeholder Participation**

As indicated by the data, the vast majority of articles involved stakeholders participation throughout the design process (65 of 88 articles; 74%), emphasizing that consistent participation is common in PD. When stakeholders are involved consistently throughout a design process – in timing and approach – it becomes much easier for practitioners to ensure that the process is equitable. More opportunities for stakeholder input at more stages of the design process will inherently amplify stakeholder voices to ensure they are considered with ample weight when design decisions are made. Even with consistent stakeholder involvement though, the equitable PD processes begin with equitable recruitment.

Often it can be difficult to find participants, with many projects relying on volunteers. This may lead to an unrepresentative group of participants and inequitable solutions, which can be avoided with activities that thoroughly recruit diverse participants. Recruitment activities such as advertising the project in the community, leveraging word of mouth (Francis, 1988), utilizing a sales pitch to encourage widespread participation (Dearden et al., 2006), and incentivizing participation encourage equitable recruitment. At its core, recruitment should be open, with attention paid to who is replying to invitations, who is participating in activities, and how participants are receptive to the design process. The methods used for recruiting participants often rely on the specifics of the project. This includes significant leg work to advertise the project for volunteers if the general public is the audience (Dalsgaard, 2012), or it could be a more targeted recruitment that only includes members of a particular organization that organized the project (Simonsen & Kensing, 1997). Regardless of the target audience for recruitment, the process must be equitable – participation must be equally accessible for all stakeholder groups that are affected by the design project; materials distributed to inform stakeholders of the opportunity to participate are clear, communicative, and inclusive; and participatory techniques used during the design process should be accessible for all participants.

### 5.4. Nuances of Participatory Techniques

Participatory Techniques that were used in the literature are significantly more nuanced than a short definition. Our research team felt that defining these techniques at a high level was valuable for a common understanding of the basic principles of each technique, but we realize that the ways in which the techniques are implemented can – and in the spirit of adaptability, should – stray from the explicit definitions in Table 5.

To offer another level of analysis that builds upon the definitions of participatory techniques, here we evaluate each technique as placed on a Spectrum of Directness, seen in Figure 4, with relation to the involvement of participants. For the purposes of this paper we will define a *direct technique* as one where participants are present and actively involved in the design activities that are a part of the technique. An *indirect technique* will be defined as one where there is not consistent, direct interaction between designers and participants.

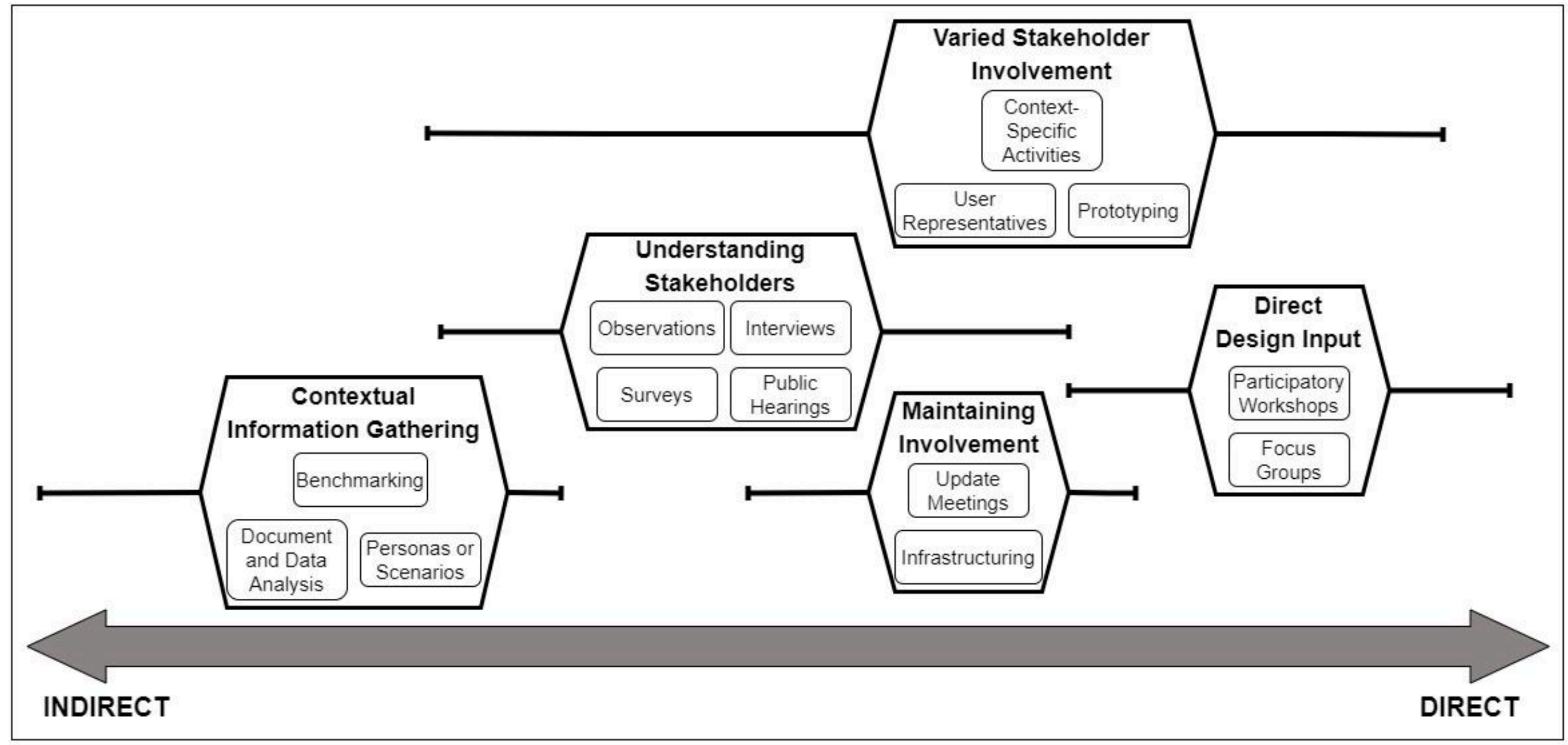

Fig. 4. Participatory techniques categorized and placed along the Spectrum of Directness.

This delineation between *direct* and *indirect* participatory techniques may seem contradictory when discussing PD. It begs the question – how can a so-called 'participatory' technique only involve stakeholders *indirectly*? The key is that a variety of multiple techniques discussed in Table 5 and displayed in Figure 4 are used to build a comprehensive and effective Participatory Design process. Some specific techniques may not be directly participatory, but they are still critical for building a foundational knowledge base and practicing empathy in a participatory process. This is supported by trends across the techniques used in the literature, with only 19 articles (22%) leveraging exclusively more *direct* methods such as Participatory Workshops or Interviews, while 65 articles (74%) used or discussed at least one *indirect* method, including Benchmarking and Surveys. 4 articles did not discuss specific participatory techniques. It is clear if 74% of the literature reviewed uses an indirect element in their Participatory Design process that these techniques are both commonly used and also necessary to the successful implementation of a PD process. An additional element of this delineation is the balance between direct and indirect techniques, to which the literature indicates a preference for direct methods with explicit participation. Here, we discuss the different categories in Figure 4, explaining where the technique is placed on the Spectrum of Directness and why it is placed there.

*Varied Stakeholder Involvement* is intentionally vague – the participatory techniques that are classified here can be used in a variety of ways that are both direct and indirect stakeholder involvement as described below. With Context-Specific Activities, User Representatives, and Prototyping, this category consists on some level of both understanding stakeholders and providing stakeholders a platform to have input on design outcomes, sometimes designing potential solutions themselves. Given this unique blend of participation, these techniques are classified as Varied Stakeholder Involvement.

*Understanding Stakeholders* has similar goals to those of ethnographic research during the design process – gain deep insights into the activities, needs, and thoughts of stakeholders. Techniques that achieve this include Observations, Interviews, Surveys, and Public Hearings. Each technique involves stakeholders contributing their insights to the designers in different ways, with the design team subsequently taking the insights to analyze and interpret them. In this way, designers learn about the stakeholders and use their input to make data-driven design decisions in alignment with stakeholder needs.

*Contextual Information Gathering* includes participatory techniques that help survey the landscape of three different areas: (1) competing or analogous design outcomes to inform successful or unsuccessful aspects of past work through Benchmarking, (2) Document and Data Analysis to analyze current work practices or trends and identify the gaps or potential needs, and (3) Personas or Scenarios to provide a contextual reference point to stakeholder needs throughout the design process. Overall, Contextual Information Gathering uses indirect methods to inform design decisions throughout the design process.

*Maintaining Stakeholder Involvement* occurs through the use of Infrastructuring or Update Meetings, when designers maintain consistent involvement with stakeholders. The level at which this is achieved differs between the two techniques, with Infrastructuring being used to drive future progress in user-led development and Update Meetings more common during the strict design process to keep users informed of progress and results. Both techniques help to keep stakeholders in the know during the design process, empowering them to contribute with the knowledge gained from these techniques.

*Direct Design Input* features two techniques – Participatory Workshops and Focus Groups – that directly involve stakeholders in insightful discussions and activities to elicit feedback and help guide the design process. These two techniques are categorized as Direct Design Input because they can go beyond information gathering at a base level, instead driving insightful, face-to-face participation during various stages of the design process to directly influence design decisions.

Even with the techniques defined in Section 4.5 and classified in this section, it is important to note that the ability to adapt is key to a successful Participatory Design process and that future applications may not fully align with the definitions or classifications in this paper. The basics of each PD technique can be taken and leveraged to best suit the specific needs of a design context, stakeholder group, timeline, or other variable – as long as it is done equitably and for the benefit of stakeholders. One simple way to begin to implement PD techniques equitably and effectively is to use a variety of techniques at a variety of stages throughout the Participatory Design process. The more techniques that are used, the more different opportunities stakeholders have to participate. If these techniques happen at various times throughout the design process, more diverse voices can be heard at different points in time. Variety in technique and timing is not a singular answer to equitable Participatory Design, but it is a first step to empowering a diverse group of stakeholders to have equitable opportunities to contribute to the design process.

### 5.5. Key Leverage Points in the Design of Participatory Design Processes

Participatory Design is inherently flexible and contextual. It must adapt to the design problem at hand to be successful. While the differences between two PD processes may appear minimal at first, a closer examination reveals that PD approaches are and should be determined based on the complex system of interrelated characteristics of the design problem. The characteristics we unpacked in this review can be a lens to support decision making about PD choices. It is essential to consider all of the "variables" or leverage points (Meadows, 2008) in a participatory process and how they interact in order to make decisions about the most appropriate PD approach in that context. Beyond the choice of the specific participatory design technique used, any researcher or practitioner creating a Participatory Design process must make decisions about several leverage points listed below and described in greater detail in Table 6.

1. Emergent vs predetermined participatory processes
2. Direct vs indirect stakeholder participation
3. Early vs late stakeholder participation
4. One time vs iterative participatory processes
5. Use of singular vs multiple participatory design techniques

Table 6. Leverage Points in a Participatory Design Process

| | *Variable* | *Advantages* | *Limitations* |
|---|---|---|---|
| 1. Emergent vs Predetermined | Emergent | Responsiveness to stakeholder needs, evolving pace of the design process, and evolving resource availability. | Source of uncertainty for the stakeholders and designers; if resources available for PD are finite and unlikely to increase, there is potential for an emergent process to exceed resource constraints. |
| | Predetermined | Source of certainty for the designers and stakeholders; predetermined processes may be looked upon favorably by funding agencies because the resource needs (space, design materials, stakeholder compensation etc.) are known. | Predetermined processes may not be able to respond to the emergence of new information or constraints during the design process; it may be difficult to onboard additional stakeholders in the midst of a predetermined process. |

| 2. Direct vs Indirect Stakeholder Participation | Direct | Stakeholders are able to directly provide input and have a say in the design process. | Direct participation of a large number of stakeholders may be difficult to schedule and requires large commitments away from the daily lives of stakeholders. |
|---|---|---|---|
| | Indirect | Allows the designers to learn about the design context and stakeholders with the investment of fewer resources. | Indirect approaches, particularly those solely relying on secondary data or limited observation risk arriving at conclusions that are not generalizable and valid. |
| 3. Early vs Late Stakeholder Participation | Early | Stakeholders involved early can influence problem formulation and shape early design ideas which often persist into the late stages of design; early participation may be less resource intensive as early stages of design typically involve low-fidelity prototyping and sketching. | Stakeholders participating early (not in the later stages of design) may not be in agreement with the evolution of their design ideas. |
| | Late | Late participation allows stakeholders to assess and test the ultimate design. | Stakeholders only participating late may not agree with the early stage design choices on which the final design is premised; they may not buy into the design. |
| 4. One Time vs Iterative Participatory Processes | One Time | One time participation may be more economical and may be more desirable for stakeholders who have limited time to commit to a PD process. | Stakeholders may feel that they do not sufficiently have a voice in the PD process and may feel like research subjects vs equal co-creators, which may also impact their trust in the designers. |
| | Iterative | An iterative process allows stakeholders to give their input across multiple stages of the design process. | Iterative processes may be resource intensive; stakeholders participating early may not be able to participate in the later stages of design if the process is prolonged and the inconsistent participation may negatively impact design outcomes. |
| 5. Singular vs Multiple Participatory Design Techniques | Singular | Singular techniques may be easier for stakeholders to learn and use; the use of singular techniques may also be less resource intensive; with the use of a singular technique it becomes possible to compare stakeholder input across the design process in a standardized manner. | Singular techniques may not be able to capture stakeholder input fully. For example, some stakeholders may be more comfortable with being interviewed vs participating in a hands-on workshop where they may not feel comfortable participating vocally, which would result in their input not |
| | | | being a part of the design process. |
| | Multiple | Multiple techniques have the advantage of being able to capture stakeholder input in many different forms and processes; stakeholders who may be less comfortable with one technique may be more comfortable with another. | It may be difficult for the same set of stakeholders to adapt to a wide range of different techniques; it may be difficult to systematically analyze data gathered from across many different techniques. |

Ignoring any of these variables risks overlooking valuable insights about how to use PD effectively. Viewing Participatory Design processes as a complex system with manageable variables is a key takeaway from this paper, guiding future research to refine and improve PD practices for more equitable outcomes. A Participatory Design process can be crafted to be equitable from the outset by managing each leverage point.

### 5.6. Rethinking Participatory Design for Equitable Outcomes

To emphasize the importance of designers actively fostering equitable processes, we explore an example from Peru, focusing on the development of informal settlements. The case illustrates how PD can be used to include typically disadvantaged stakeholders, but also shows how it can inadvertently expand the power of dominant groups (Frediani, 2016). Power imbalances in PD processes can manifest in several ways: the design of the process to benefit a particular group through the manipulation of the leverage points discussed in the previous section, the dilution of diverse stakeholder needs into overly generalized findings, the suppression of quieter voices in large-scale contexts, or overly constrained solutions that limit new learning opportunities. These scenarios demonstrate that simply involving stakeholders in design does not guarantee an equitable process or outcome.

The power dynamics in these projects raise important questions about democracy in design and how PD can empower stakeholders without reinforcing existing power disparities. When implemented well, PD can have far-reaching benefits across design domains that require flexible processes and adaptable solutions to wicked problems (Rittel & Webber, 1973) – problems that affect large populations and may lead to solutions that harm minoritized communities when all stakeholders are not provided a voice in the design process. The first step in empowering stakeholders is honing the five leverage points discussed in the previous section during the design of the Participatory Design process: (1) emergent vs. predetermined design processes; (2) direct vs indirect stakeholder participation; (3) early vs late stakeholder participation; (4) one time vs iterative participation; and (5) singular vs multiple PD techniques. More equitable design outcomes are the product of equitable design processes that are crafted to be equitable from the very outset using these leverage points.

Simply involving stakeholders does not automatically lead to an equitable design process or just outcomes. Other practitioners have called for a rethinking of PD as a meta-methodology, moving beyond traditional practices to a more radical approach – Radical Participatory Design – which critically examines power imbalances between practitioners and stakeholders (Udoewa, 2022). This discussion is essential to challenge traditional notions of equity in design. Engaging with stakeholders and users is a first step towards a future of design that is participatory, effective, and equitable.

### 5.7. Future Work

Further research into the nuanced aspects of equity, empowerment, and their role in defining successful outcomes in Participatory Design is essential as the engineering design community works towards a future of equitable design practices. Practitioners are already exploring this area, focusing on integrating compassion in Participatory Design processes through practitioner reflections that emphasize stakeholder's dignity, empowerment, and security (Seshadri et al., 2019) as well as redesigning design processes to include and empower novices and non-designers (Efeoğlu & Møller, 2023). These efforts contribute to bridging the gap in applied equitable Participatory Design research, which is vital for helping engineers create more equitable processes and outcomes.

As discussed in this paper, future work is needed in exploring Participatory Design processes as decisions within a complex system, investigating the leverage points from the previous section. Key questions include which variables contribute most to stakeholder satisfaction with the outcome or with their involvement in the process. Focusing on how each variable influences equity within the process can guide practitioners in structuring PD processes that offer more value to stakeholders and provide clearer direction for designers.

Further research should also examine the effectiveness of specific participatory techniques. Research should also explore the potential negative effects of certain techniques, such as whether they cause conflict between stakeholder groups and how they can be modified to reduce such tensions. As techniques are central to Participatory Design, understanding their impact will help practitioners develop best practices to empower stakeholders.

## 6. CONCLUSION

With this paper, we first examined what constitutes Participatory Design in the field by collating design literature from an array of sources and analyzing it to address a twin desire across design disciplines – engineering design in particular – to remedy inequitable design by engaging directly with stakeholders and users. This paper shows what PD looks like in practice, drawing from past PD processes to improve its future use for equitable processes and outcomes.

Our team determined multiple salient trends in the literature. A majority of the Participatory Design literature discussed specific case studies. The contexts in which Participatory Design were applied showed a majority of applications with intangible systems – with an overwhelming majority occurring in real-world projects. We saw that the most successful Participatory Design processes put in significant foundational work to recruit stakeholders, with methods tailored to recruit those who best represent the stakeholder group. Once recruited, stakeholders participated throughout the design process in a significant majority of the literature, pointing to consistency being a key for stakeholder participation. This showed us that consistently involving stakeholders leads to a democratic design process, although said process must begin with equitable recruitment.

Once recruited, 14 distinct participatory techniques – described in Section 4.5 – were used to engage stakeholders in Participatory Design processes. A deeper analysis of these techniques manifested Figure 4 in Section 5.4 – the Spectrum of Directness. This spectrum allows for fluidity and flexibility in our definition of participatory techniques, demonstrating the oftentimes ambiguous nature of PD processes – and transitively the techniques practitioners use – that emphasizes the crucial nature of adaptability in PD.

Analyzing Participatory Design as a complex system, we determined five key leverage points, or variables in the design of the process: (1) emergent vs. predetermined design processes; (2) direct vs indirect stakeholder participation; (3) early vs late stakeholder participation; (4) one time vs iterative participation; and (5) singular vs multiple PD techniques. By managing these variables at the outset of the design process, practitioners can tune their process to a specific design context, while ensuring that the process itself is designed equitably. Equitable design outcomes require equitable design processes, and the five key variables are the first step towards designing an equitable process.

In addition to the findings described above, we would like to emphasize – any design process at its core must embody equity. As such, acknowledging the power dynamics in a Participatory Design process and making every attempt to mitigate undesirable dynamics are paramount. Designers must maintain their focus on empowerment, especially when the line between empowerment and abuse of power in Participatory Design is a close one.

**ACKNOWLEDGMENTS**

Thank you to Nate Piersma, who participated in the review of literature for data collection. We gratefully acknowledge support for this research from a Fastest Path to Zero Mini Grant and a Catalyst Grant from the Graham Sustainability Institute.